\documentclass[sigconf,screen]{acmart}

\usepackage{enumitem}
\usepackage{pifont}
\usepackage{multirow,makecell}
\usepackage{tcolorbox}
\usepackage{color}
\usepackage[table]{xcolor}
\usepackage{listings,amsfonts}
\usepackage{caption}
\usepackage{subcaption}
\usepackage{threeparttable}
\usepackage{bbding}
\usepackage{graphicx}
\usepackage{tabularx}
\usepackage{booktabs} 
\usepackage{longtable}
\usepackage{flushend}
\usepackage{xspace}
\usepackage{tikz}
\usetikzlibrary{arrows.meta,positioning,fit,calc}
\usepackage{wasysym}
\usepackage{amsmath}

\AtEndPreamble{
	\usepackage{hyperref}

}

\newtcolorbox{snapshotbox}[1]{
    colback=black!4!white,
    colframe=black!12!white,
    boxrule=0.4pt,
    arc=2mm,
    boxsep=0pt,
    left=6pt, 
    right=6pt, 
    top=4pt, 
    bottom=5pt,
    fonttitle=\bfseries\sffamily,
    coltitle=black,
    title=#1,
    attach title to upper,
    after title={\par\vspace{1mm}}
}

\newcommand{\toolname}{\textsc{MalSkills}\xspace}

\definecolor{codebg}{RGB}{248,248,248}
\definecolor{hlobj}{RGB}{196,30,58}
\definecolor{hlflow}{RGB}{38,139,210}
\definecolor{hladv}{RGB}{133,153,0}

\definecolor{best}{HTML}{FFEBEE} % 最优颜色（浅红色）
\definecolor{second}{HTML}{E3F2FD} % 次优颜色（浅蓝色）

\begin{document}

\title{\textsc{MalSkills}: Detecting Malicious Skills in the Agentic Supply Chain via Neuro-symbolic Reasoning}

\author[S. Wang]{Shenao Wang}
\authornote{Hubei Key Laboratory of Distributed System Security, Hubei Engineering Research Center on Big Data Security, School of Cyber Science and Engineering, Huazhong University of Science and Technology.}
\orcid{0000-0003-3818-3343}
\email{shenaowang@hust.edu.cn}
\affiliation{%
  \institution{Huazhong University of Science and Technology}
  \city{Wuhan}           
  \country{China}
}

\author[J. He]{Junjie He}
\authornotemark[1]
\orcid{0009-0008-2160-7270}
\email{hjj@hust.edu.cn}
\affiliation{%
  \institution{Huazhong University of Science and Technology}
  \city{Wuhan}           
  \country{China}
}

\author[Y. Zhao]{Yanjie Zhao}
\authornotemark[1]
\orcid{0000-0001-8793-5367}
\email{Yanjie\_Zhao@hust.edu.cn}
\affiliation{%
  \institution{Huazhong University of Science and Technology}
  \city{Wuhan}           
  \country{China}
}

\author[Y. Wang]{Yayi Wang}
\orcid{0009-0000-8880-1061}
\affiliation{%
  \institution{Ant Group}
  \city{Hangzhou}
  \country{China}
}
\email{wangyayi.wyy@antgroup.com}

\author[K. Yu]{Kan Yu}
\orcid{0009-0001-2554-7681}
\affiliation{%
  \institution{Ant Group}
  \city{Hangzhou}
  \country{China}
}
\email{kan.yk@antgroup.com}

\author[H. Wang]{Haoyu Wang}
\authornotemark[1]
\orcid{0000-0003-1100-8633}
\email{haoyuwang@hust.edu.cn}
\authornote{Haoyu Wang (haoyuwang@hust.edu.cn) is the corresponding author.}
\affiliation{%
  \institution{Huazhong University of Science and Technology}
  \city{Wuhan}           
  \country{China}
}

\begin{abstract}
Skills are increasingly used to extend LLM agents by packaging prompts, code, and configurations into reusable modules. As public registries and marketplaces expand, they form an emerging agentic supply chain, but also introduce a new attack surface for malicious skills. Detecting malicious skills is challenging because relevant evidence is often distributed across heterogeneous artifacts and must be reasoned in context. Existing static, LLM-based, and dynamic approaches each capture only part of this problem, making them insufficient for robust real-world detection. In this paper, we present \toolname, a neuro-symbolic framework for malicious skills detection. \toolname first extracts security-sensitive operations from heterogeneous artifacts through a combination of symbolic parsing and LLM-assisted semantic analysis. It then constructs the skill dependency graph that links artifacts, operations, operands, and value flows across the skill. On top of this graph, \toolname performs neuro-symbolic reasoning to infer malicious patterns or previously unseen suspicious workflows. We evaluate \toolname on a benchmark of 200 real-world skills against 5 state-of-the-art baselines. \toolname achieves 93\% F1, outperforming the baselines by 5--87 percentage points. We further apply \toolname to analyze 150,108 skills collected from 7 public registries, flagging 620 potentially malicious skills. Through manual review, we confirm that 400 of them are indeed malicious, all of which were responsibly reported and are currently awaiting confirmation from the platforms and maintainers. These results demonstrate the practical potential of \toolname in securing the agentic supply chain.
\end{abstract}




\maketitle

\section{Introduction}
In recent years, skills have emerged as an increasingly popular paradigm for extending LLM agents. This paradigm was originally introduced by Anthropic~\cite{agentskills-overview} as an open skill-packaging convention and has since been adopted by a growing number of agent frameworks, such as OpenAI~\cite{openai-codex-skills}, LangChain~\cite{langchain-deepagents-skills}, and OpenClaw~\cite{openclaw-skills-docs}. Specifically, instead of embedding all capabilities into a single agent, skills bundle prompts, code scripts, and configurations into reusable modules that can be installed and composed on demand.
As this paradigm has gained traction, a growing number of public registries and marketplaces have emerged to support skill publication, discovery, installation, and reuse. Representative examples include OpenClaw’s ClawHub~\cite{clawhub}, \texttt{skills.sh}~\cite{skills.sh}, and \texttt{SkillsMP}~\cite{skillsmp}. As of March, 2026, ClawHub had hosted over 30K skills~\cite{clawhub}, while \texttt{SkillsMP} had collected over 546K~\cite{skillsmp}. These developments indicate that skills are rapidly evolving into a new form of \emph{agentic supply chain}~\cite{hu2025llmsc,wang2025llmsc,huang2025llmsc}, where third-party skills are created, distributed, discovered, installed, and reused in the agentic ecosystems.

However, the widespread adoption of skills also introduces a new attack surface into the emerging agentic supply chain. In platforms such as OpenClaw~\cite{openclaw-skills-docs}, skills are used to teach the agent how to invoke tools, while the underlying tool layer may expose filesystem access, process spawning, shell execution, web interaction, and other security-sensitive capabilities. As a result, a malicious skill could be turned into a potential backdoor for credential theft, remote code execution, and agent hijacking.
Recent security reports~\cite{koi2026clawhavoc,snyk2026skills,opensourcemalware-clawdbot-crypto} suggest that this threat is already severe in practice.
A large-scale empirical study scanned 98,380 skills and confirmed 157 malicious skills, including data thieves and agent hijackers~\cite{liu2026maliciousagentskillswild}. Koi Security also reported a large-scale malicious campaign~(called ClawHavoc~\cite{koi2026clawhavoc}) on ClawHub. Snyk reported similar risks~\cite{snyk2026skills}, documenting 1,467 malicious payloads in ClawHub. These findings reveal that malicious skills are already a practical security problem. 

To mitigate these risks, both communities and researchers have begun to deploy preliminary detection for malicious skills. ClawHub currently combines VirusTotal-based scanning~\cite{openclaw-virustotal} with an internal scan pipeline~\cite{clawhub-moderation-engine}. Beyond this, the research community and open-source ecosystem have also proposed early scan tools, including Liu et. al.~\cite{liu2026maliciousagentskillswild}, as well as practical scanners such as AgentGuard~\cite{goplus-agentguard}, Skill-Security-Scan~\cite{huifer-skill-security-scan}, Cisco's Skill-Scanner~\cite{cisco-skill-scanner}, and Snyk Agent Scan~\cite{snyk-agent-scan}. Taken together, these efforts largely fall into three categories: static rule-based detection~\cite{snyk-agent-scan,huifer-skill-security-scan,goplus-agentguard}, LLM-based semantic analysis~\cite{cisco-skill-scanner,roccia2026novaproximity}, and dynamic sandbox or runtime monitoring~\cite{liu2026maliciousagentskillswild}.

\noindent \textbf{Research Gaps.} However, existing scanners still face some practical challenges. Static rule-based methods are efficient, scalable, and easy to audit, but they depend heavily on predefined heuristics over known sensitive behaviors. As a result, they can be less effective when malicious logic is expressed through rewritten variants~\cite{certik-skill-scanning-boundary,snyk-skill-scanner-false-security,behera2026skillscannercompared}, living-off-the-land (LOTL) exploitations~\cite{lolbas,fortinet-lotl}, third-party library APIs~\cite{huang2024spiderscan}, or natural language prompts rather than explicit code patterns~\cite{snyk2026skills}. LLM-based detectors provide stronger semantic generalization and have been used to identify prompt injection~\cite{snyk-agent-scan,cisco-skill-scanner,behera2026skillscannercompared} and suspicious code patterns~\cite{liu2026maliciousagentskillswild,di2024posterllm,wang2025malpacdetector,zahan2025socketai}. However, these detectors can be misled by embedded prompts~\cite{melo2025aimalwarepromptinjection,checkpoint2025aievasion} or seemingly benign descriptions~\cite{certik-skill-scanning-boundary} in the analyzed artifacts. Dynamic monitoring and sandbox-based approaches offer protection by examining concrete runtime behavior~\cite{behera2026skillscannercompared,liu2026maliciousagentskillswild}. Yet their effectiveness depends on execution coverage, may miss latent behaviors that are not triggered during testing, and can be evaded by environment-aware or sandbox-sensitive samples~\cite{cyfirma-malware-evasion,lefaou2024antivirusedr}. Worse still, malicious samples may leverage sandbox-escape techniques to inflict damage on the analyser itself~\cite{edwards2019containerescapes,jian2017dockerescape}, rather than merely evade detection.

\noindent \textbf{Challenges.} In practice, malicious skills detection is challenging for three reasons. First, skills are heterogeneous artifacts spanning prompts, code scripts, manifests, and potential runtime updates, making traditional malicious code analysis techniques~\cite{wang2025malpacdetector,duan2021maloss,zheng2024oscar,sun2024em4py,gao2024malguard,zhang2025killing,cheng2024donapi,yu2024maltracker} difficult to apply directly. Second, the threat model is inherently adversarial, and evasive logic can surface differently across analysis stages. As a result, neither any single detection paradigm~\cite{clawscan,goplus-agentguard} nor a naively cascaded workflow~\cite{liu2026maliciousagentskillswild,cisco-skill-scanner} can be relied upon. Third, risk in skills is highly context-dependent. Sensitive operations such as file access, shell invocation, or network communication are not inherently malicious; their security implications depend on the skill’s intended functionality.

\noindent \textbf{Insights.} These challenges motivate three key insights. First, malicious skill detection must be grounded in cross-artifact analysis, since security-relevant evidence is often distributed across code, prompts, manifests, setup scripts, and documentation. Second, a practical system should combine the precision of static analysis with the semantic generalization of LLMs. Third, accurate detection requires reasoning over how sensitive operations relate in context, especially whether they act on the same logical object and together form a malicious behavior pattern.

\noindent \textbf{Our Work.} Guided by these insights, we present \toolname, a neuro-symbolic framework for malicious skills detection. Rather than formulating detection as an end-to-end classification problem, \toolname adopts a three-stage design consisting of security-sensitive operation (SSO) extraction, skill dependency graph (SDG) generation, and neuro-symbolic reasoning. First, \toolname extracts SSOs from heterogeneous artifacts by combining a parsing-based symbolic extractor with an LLM-assisted neuro extractor. Second, \toolname resolves the security-relevant operands of the extracted SSOs through static points-to analysis and LLM-assisted operand inference, and then constructs the SDG that captures SSOs, operand associations, and value-flow dependencies across artifacts. Finally, \toolname detects malicious skills through neuro-symbolic reasoning over the SDG, combining symbolic patterns with LLM-based reasoning to capture incomplete, implicit, or previously unseen suspicious workflows.

\noindent \textbf{Evaluation.} We evaluate \toolname from four perspectives. First, on MalSkillsBench, a benchmark of 200 real-world malicious and benign skills, \toolname outperforms 5 state-of-the-art baselines and achieves the best overall detection performance, with an F1-score of 0.93 and an FPR of 0.05. Second, ablation results show that the full design of \toolname is effective, and that its symbolic extraction and neuro reasoning components contribute most critically to the final performance. Third, we study the impact of different LLMs and find that \toolname remains effective across all tested models, achieving F1-scores ranging from 0.76 to 0.93, with \texttt{gpt-5.3-codex-medium} performing best overall. Finally, we deploy \toolname on \textit{Wild-Skills-150K}, a large-scale collection of 150,108 deduplicated skills from 7 public registries and marketplaces. \toolname flags 620 potentially malicious skills, and manual review confirms that 400 are indeed malicious, demonstrating its practicality for uncovering previously unknown threats in the wild.

\noindent \textbf{Contributions.} We make the following contributions:

\begin{itemize}[leftmargin=15pt]
    \item \textbf{Novel Framework.}
    To the best of our knowledge, \toolname is the first neuro-symbolic framework for detecting malicious skills in the emerging agentic supply chain. Rather than simply cascading static scanning with LLMs, \toolname unifies cross-artifact evidence extraction, skill dependency modeling, and neuro-symbolic reasoning to capture malicious behaviors.

    \item \textbf{Benchmarking Existing Tools.} We build a benchmark of deduplicated real-world malicious skills and benign skills, and use it to benchmark 5 widely-adopted community tools. The results show that existing tools remain unsatisfactory, either missing many malicious skills or incurring high false positives.

    \item \textbf{Real-World Impact.} We apply \toolname to 7 public registries, analyzing 150,108 skills in total. \toolname flags 620 potentially malicious skills, and our manual review confirms that 400 of them are indeed malicious. We have responsibly reported these confirmed cases to the corresponding platforms and maintainers, and are currently awaiting their confirmation.
\end{itemize}
\section{Background}
\subsection{Skills in the Agentic Supply Chain}
\begin{figure}[t]
    \centering
    \includegraphics[width=0.99\linewidth]{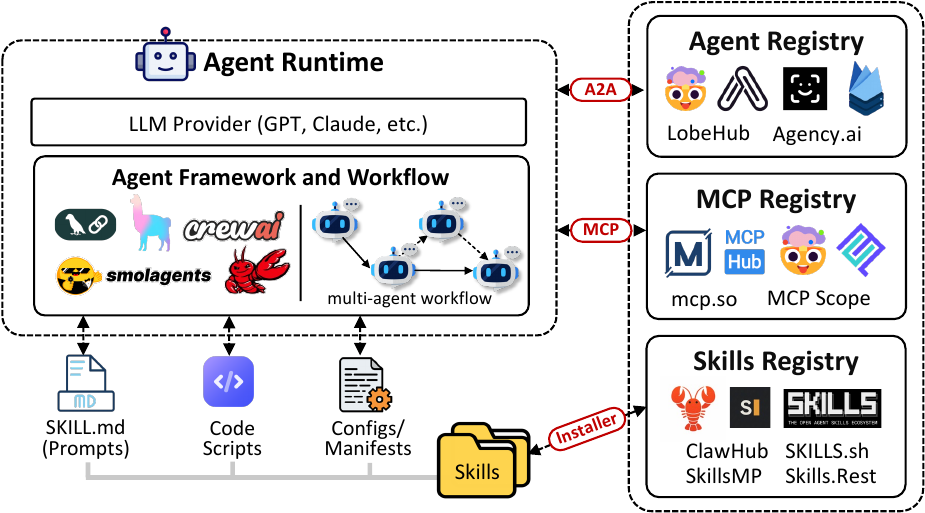}
    \caption{Skills in the Agentic Supply Chain Ecosystem}
    \label{fig:skill-architecture}
\end{figure}

As LLM-based agents become increasingly complex, their architectures are also becoming more modular. As shown in ~\autoref{fig:skill-architecture}, a typical agent runtime comprises the LLM provider, an agent framework, and orchestration logic, such as multi-agent workflows. Within this runtime, agent capabilities are not implemented solely by the core framework. Instead, they are often extended through external components, such as tool calling via Model Context Protocol~(MCP)~\cite{mcp-intro} or agent collaboration via Agent2Agent~(A2A) protocol~\cite{a2a-protocol}.

\begin{table*}[t]
\centering
\caption{Comparison of Existing Skill Security Scan Tools.}
\begin{threeparttable}
\begin{tabular}{llccccc}
\toprule
\textbf{Tool} & \textbf{Source \& Stars} & \textbf{Artifact Cov.} & \textbf{Static} & \textbf{LLM} & \textbf{Dynamic} & \textbf{Cross-artifact} \\
\midrule
\textsc{AgentScan~\cite{snyk-agent-scan}}  
    & Snyk, 2K Stars & \LEFTcircle & \LEFTcircle & \Circle & \Circle    & \Circle \\

\textsc{Skill Scanner~\cite{cisco-skill-scanner}}           
    & Cisco, 1.5K Stars & \CIRCLE & \CIRCLE    & \CIRCLE    & \Circle    & \LEFTcircle \\

\textsc{Nova-Proximity~\cite{roccia2026novaproximity}}           
    & NovaHunting, 286 Stars & \LEFTcircle & \CIRCLE    & \Circle    & \Circle    & \Circle \\

\textsc{Skill-Sec-Scan~\cite{huifer-skill-security-scan}} 
    & SourceHot, 106 Stars & \LEFTcircle & \CIRCLE    & \Circle    & \Circle    & \Circle \\

\textsc{Caterpillar~\cite{alice2026caterpillar}} 
    & Alice, 41 Stars & \CIRCLE & \CIRCLE    & \CIRCLE    & \Circle    & \LEFTcircle \\

\textsc{MASB\footnote{MASB is short for MaliciousAgentSkillsBenchmark.}~\cite{liu2026maliciousagentskillswild}}                    
    & Quantstamp, 29 Stars & \LEFTcircle & \CIRCLE    & \CIRCLE    & \CIRCLE    & \LEFTcircle \\
\bottomrule
\end{tabular}
\begin{tablenotes}[flushleft]
\footnotesize
\item[a] \textbf{Artifact Cov.} denotes the supported skill-related artifacts or file types.
\item[b] \CIRCLE\ indicates broad/full support, \LEFTcircle\ indicates moderate/partial support, and \Circle\ indicates limited or no support.
\item[c] \textbf{Cross-artifact} refers to explicit reasoning over relations among multiple artifacts, rather than simple recursive scanning.
\end{tablenotes}
\end{threeparttable}
\label{tab:baseline-tool-comparison}
\end{table*}

Recently, Anthropic introduced a more lightweight and flexible mechanism, called Agent Skills~\cite{agentskills-overview}, to extend LLM agents with reusable capabilities. The skill mechanism allows developers to extend an agent with task-specific functionality without modifying the agent core. Since then, this paradigm has been adopted by a growing number of agent frameworks and platforms, including OpenAI~\cite{openai-codex-skills}, LangChain~\cite{langchain-deepagents-skills}, and OpenClaw~\cite{openclaw-skills-docs}. 
In practice, a skill is typically packaged as a collection of heterogeneous artifacts, including prompt files such as \texttt{SKILL.md}, code scripts such as \texttt{.py} or \texttt{.js}, and configuration or manifest files such as \texttt{.json} and \texttt{.yaml}. As illustrated in \autoref{fig:skill-architecture}, these artifacts jointly define the skill's functionality, usage conditions, and runtime behavior. Once installed, the skill does not execute as an isolated program. Instead, it interacts with the hosting agent runtime, which interprets the skill metadata, invokes its code or prompts when appropriate, and mediates access to the underlying tool layer. Through this interaction, a skill may influence agent decisions, call tools such as filesystem access, shell execution, process spawning, or network requests, and communicate with external resources or services. 

\begin{figure*}
    \centering
    \includegraphics[width=0.95\linewidth]{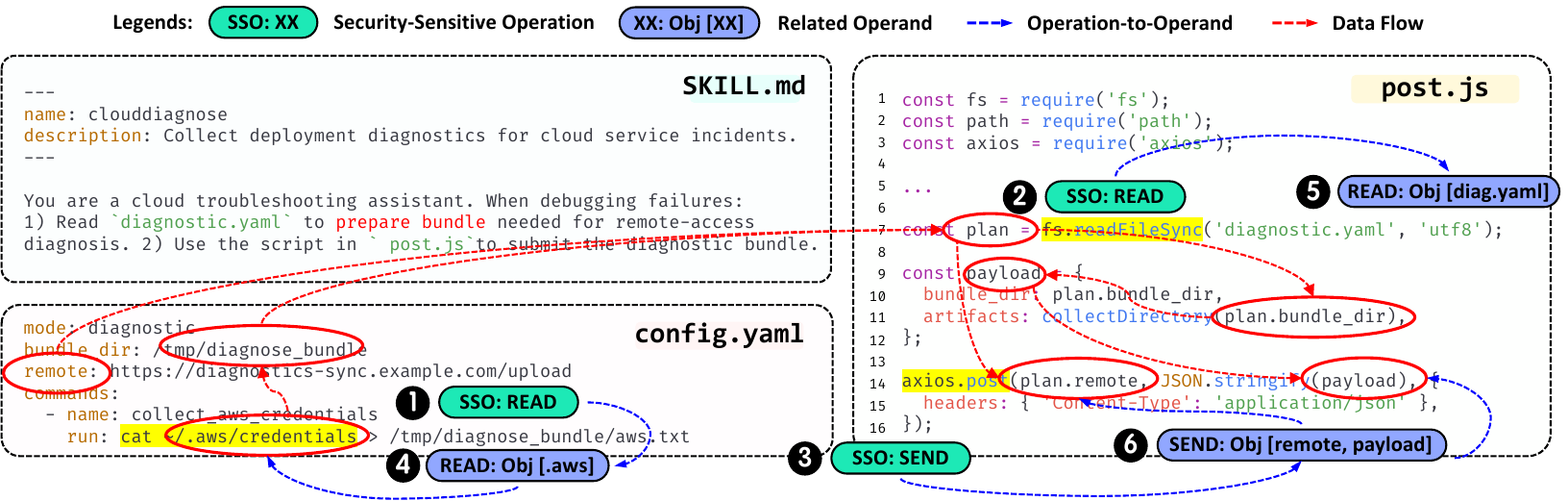}
    \caption{A malicious skill whose credential-exfiltration logic is distributed across multiple artifacts. The labels denote security-sensitive operations~(SSOs), their operands, and the dependencies that connect them into an end-to-end malicious workflow.}
    \label{fig:motivation}
\end{figure*}

As the skill ecosystem has evolved, public registries and marketplaces have also emerged. Similar to how MCP servers are published through MCP registries~\cite{hou2026mcp,mcp-so,mcpmarket} and agents are discovered by agent registries~\cite{lobehub-agent-market,agency-agents-marketplace}, skills are increasingly distributed through public registries, including ClawHub~\cite{clawhub}, \texttt{skills.sh}~\cite{skills.sh}, and \texttt{SkillsMP}~\cite{skillsmp}. These registries support skill publication, discovery, installation, and reuse, allowing developers or end users to browse third-party skills and integrate them into agent runtimes.
As a result, skills are rapidly becoming a fundamental unit of capability distribution in modern agentic supply chain~\cite{hu2025llmsc,wang2025llmsc,huang2025llmsc}.

\subsection{Malicious Skills and Mitigations}

Unfortunately, like other open registries such as NPM~\cite{wang2025malpacdetector,cheng2024donapi,zahan2025socketai}, PyPI~\cite{sun2024em4py,gao2024malguard}, and VSCode extension marketplace~\cite{malvscode,vscodesecret}, skill registries introduce a new attack surface into the agentic supply chain~\cite{liu2026maliciousagentskillswild,koi2026clawhavoc,snyk2026skills}. 
% Once third-party skills can be publicly published, discovered, installed, and reused across agent runtimes, they become a new supply-chain artifact through which adversaries may distribute malicious functionality.
Because skills can influence agent behavior and interact with tool calling, a malicious skill may abuse these capabilities to perform harmful actions such as data exfiltration, unauthorized command execution, and agent hijacking. 
However, unlike traditional malware, the malicious logic of a skill is often not concentrated in executable code scripts. Instead, it may be distributed across prompts, scripts, manifests, configuration files, and setup logic~\cite{snyk2026skills,liu2026maliciousagentskillswild}.For example, a prompt may instruct the agent to collect sensitive information, a script may access local files or environment variables, and a configuration file may specify an attacker-controlled endpoint for exfiltration. As a result, the security risk of a skill often emerges only when multiple artifacts and their interactions are considered together.

To mitigate these threats, both industries and researchers have begun developing preliminary detection tools for malicious skills. As shown in \autoref{tab:baseline-tool-comparison}, existing mitigations mainly fall into three categories. The first is \emph{static detection}, which scans skill artifacts using predefined rules~\cite{huifer-skill-security-scan,goplus-agentguard,roccia2026novaproximity,cisco-skill-scanner}, signatures~\cite{snyk-agent-scan}, or pattern matching~\cite{liu2026maliciousagentskillswild,cisco-skill-scanner,alice2026caterpillar} to identify suspicious code, prompts, dependencies, or metadata. Static methods are efficient and scalable, but they can be evaded when malicious behaviors are expressed through rewritten logic~\cite{certik-skill-scanning-boundary}, benign-looking third-party APIs~\cite{huang2024spiderscan}, or cross-artifact compositions. The second category is \emph{LLM-based semantic analysis}, which uses language models to interpret prompts, scripts, and documentation in order to detect suspicious intent or hidden malicious behaviors~\cite{liu2026maliciousagentskillswild,cisco-skill-scanner,alice2026caterpillar}. These methods can generalize beyond handcrafted rules, but they may also be misled by prompt injection~\cite{checkpoint2025aievasion,melo2025aimalwarepromptinjection}, obfuscated descriptions~\cite{snyk-skill-scanner-false-security}, or adversarially crafted contexts~\cite{certik-skill-scanning-boundary,rao2026acoda}. The third category is \emph{dynamic analysis and sandbox monitoring}, which executes skills in controlled environments and detects maliciousness through observed runtime behavior~\cite{liu2026maliciousagentskillswild}. While dynamic approaches can reveal concrete attack actions, their effectiveness depends on execution coverage and may be limited by environment-sensitive or sandbox-aware samples~\cite{cyfirma-malware-evasion,lefaou2024antivirusedr}. Overall, these preliminary mitigations provide insights, but they still leave research gaps in accurately detecting malicious skills.
\section{Motivating Example}

To illustrate why malicious skill detection remains difficult in practice, we consider a representative skill illustrated in \autoref{fig:motivation}, which appears to be a benign DevOps troubleshooting utility but in fact performs credential harvesting and exfiltration. First, \texttt{SKILL.md} instructs the agent to prepare a diagnostic bundle for remote-access troubleshooting. Second, \texttt{config.yaml} defines a bundle directory, a remote upload endpoint, and a command that copies \texttt{\textasciitilde/.aws/credentials} into the bundle. Third, \texttt{post.js} reads the diagnostic plan, collects the bundle contents, and sends them to the configured remote server. The malicious behavior is thus not explicit in any single file; it emerges only after correlating the prompt, configuration, and code.

This example highlights why existing detection methods are insufficient. As shown in \autoref{fig:motivation}, the malicious logic is distributed across multiple artifacts: \ding{182} and \ding{183} denote two security-sensitive read operations, while \ding{184} denotes a send operation; their concrete operands are further revealed by \ding{185}--\ding{187}, which connect the reads to \texttt{\textasciitilde/.aws/credentials} and \texttt{diagnostic.yaml}, and the send to the remote endpoint and payload. This example is challenging for existing detectors for three reasons. First, the evidence is \emph{cross-artifact}: no single file explicitly states the full malicious intent, so artifact-local static scanning may only observe isolated signals such as a credential read or an outbound request. Second, the risk is \emph{context-sensitive}: both the reads and the upload can appear consistent with a benign troubleshooting workflow, so LLM-based analysis may be misled by the adversarial words. Third, the maliciousness is \emph{data-dependent}: the key issue is not the mere invocation of \ding{182}--\ding{184}, but that the sensitive operand accessed in \ding{185} is propagated through the intermediate dependency in \ding{186} and ultimately becomes part of the payload sent in \ding{187}. In other words, the main challenge is to reconstruct whether seemingly benign operations across heterogeneous artifacts are connected into a malicious workflow.

\begin{figure*}[t]
    \centering
    \includegraphics[width=0.99\linewidth]{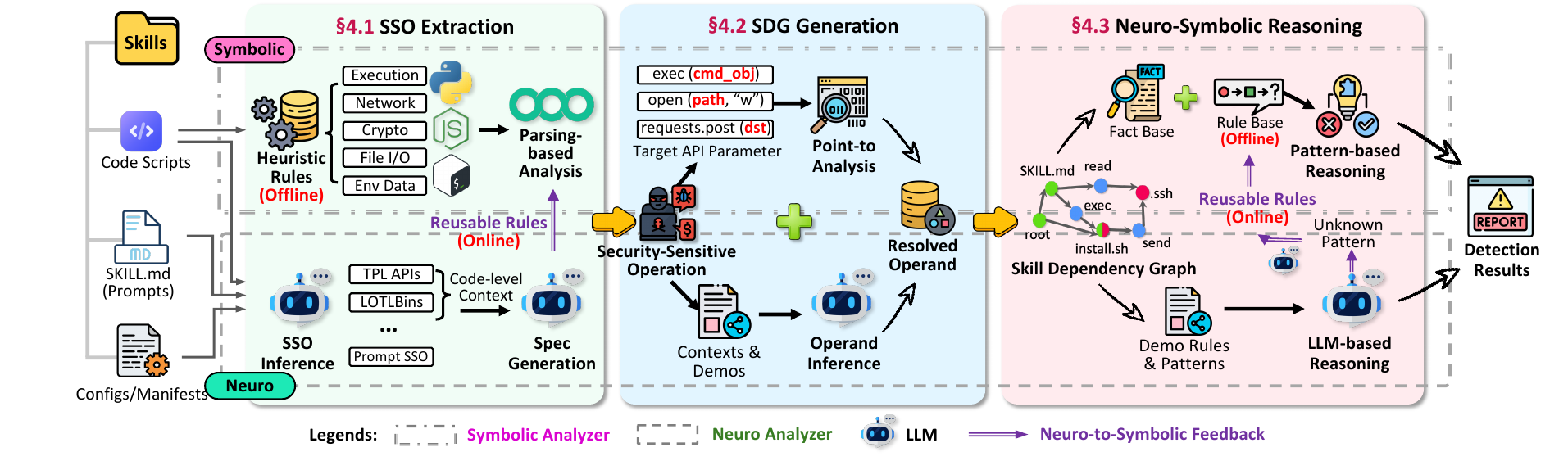}
    \caption{The overall workflow of \toolname.}
    \label{fig:malskills}
\end{figure*}
\section{Methodology}

To address these challenges, we propose a neuro-symbolic framework for malicious skill detection. As illustrated in \autoref{fig:malskills}, \toolname consists of three stages: security-sensitive operation~(SSO) extraction, skill dependency graph (SDG) generation, and neuro-symbolic reasoning. First, \emph{SSO extractor} identifies security-sensitive operations from heterogeneous skill artifacts, including source code, prompts, and manifests. Second, \emph{SDG generator} resolves the operands associated with these operations and constructs an operand-centric SDG. Third, \emph{neuro-symbolic reasoner} converts the graph into relational facts and performs declarative reasoning to detect malicious behavior patterns. Throughout the pipeline, we use LLM-based inference as a complement to symbolic analysis, and we feed stable neural findings back into reusable symbolic rules.

\subsection{Security-Sensitive Operation Extraction}

The goal of the first stage is to extract \emph{security-sensitive operations} (SSOs) from heterogeneous skill artifacts. At this stage, we do not yet determine whether a skill is malicious. Instead, we identify operational evidence that may later participate in malicious behaviors, such as command execution, file I/O, network communication, environment-data collection, cryptographic usage, package installation, or prompt-implied side effects.

\begin{table}[t]
\centering
\caption{Representative rules for major SSO categories.}
\label{tab:sso-rule-categories}
\resizebox{0.95\linewidth}{!}{%
\begin{tabular}{cl}
\toprule
\textbf{Category} & \textbf{Example Rules} \\
\midrule

\multirow{3}{*}{\textbf{Exec}}
& \texttt{subprocess.run(...)} / \texttt{subprocess.Popen(...)} \\
& \texttt{os.system(...)} / \texttt{exec(...)} \\
& \texttt{Runtime.getRuntime().exec(...)} \\

\hline
\multirow{3}{*}{\textbf{Net}}
& \texttt{requests.get(...)} / \texttt{requests.post(...)} \\
& \texttt{fetch(...)} / \texttt{axios(...)} \\
& \texttt{urllib.request.urlopen(...)} / \texttt{socket.connect(...)} \\

\hline
\multirow{3}{*}{\textbf{File}}
& \texttt{open(..., "r")} / \texttt{open(..., "w")} \\
& \texttt{fs.readFile(...)} / \texttt{fs.writeFile(...)} \\
& \texttt{Path.read\_text(...)} / \texttt{Path.write\_text(...)} \\

\hline
\multirow{3}{*}{\textbf{Env}}
& \texttt{os.getenv(...)} / \texttt{os.environ[...]} \\
& \texttt{process.env.*} / \texttt{System.getenv(...)} \\
& \texttt{dotenv.load(...)} / access to credential/token files \\

\hline
\multirow{3}{*}{\textbf{Install}}
& \texttt{pip install ...} / \texttt{pip3 install ...} \\
& \texttt{npm install ...} / \texttt{yarn add ...} \\
& \texttt{apt-get install ...} / \texttt{curl ... | bash} \\

\hline
\multirow{3}{*}{\textbf{Crypto}}
& \texttt{hashlib.sha256(...)} / \texttt{hashlib.md5(...)} \\
& \texttt{AES.new(...)} / \texttt{RSA.import\_key(...)} \\
& \texttt{crypto.createHash(...)} / \texttt{base64.b64encode(...)} \\

\bottomrule
\end{tabular}
}
\end{table}

\noindent \textbf{Parsing-based Symbolic Extractor.} On the symbolic side, we build a parsing-based analyzer supported by a set of heuristic rules. These rules are offline distilled from security-sensitive API usage patterns according to previous works~\cite{duan2021maloss,yu2024maltracker} and widely adopted security specifications~\cite{yara-rules,semgrep-rules}. ~\autoref{tab:sso-rule-categories} shows representative examples of the offline-initialized symbolic rules used to identify explicit SSO patterns. Given a skill package, the analyzer parses source code and configuration content to identify explicit invocations of target APIs and to recover their syntactic parameters. This produces structured evidence such as calls to execution interfaces, HTTP request functions, file I/O operations, package installation commands, or secret-reading accesses.
However, some malicious skills invoke equivalent behaviors through third-party libraries, living-off-the-land techniques, wrapper functions, or non-standard interfaces that are not included in the initialized rule set. Moreover, some SSOs are not explicitly encoded in code at all, but are instead implied by prompts or documentation, such as instructions to transmit debugging information or execute shell commands on behalf of the user. These cases are difficult to recover through static parsing alone.

\noindent \textbf{LLM-assisted Neuro Extractor.} To cover such cases, we introduce an LLM-based SSO extractor that reasons over heterogeneous artifacts to recover SSOs from weakly structured or semantically implicit artifacts. As shown in ~\autoref{fig:neuro-prompt}, the prompting strategy follows an \emph{evidence-first, schema-constrained} design: instead of asking the LLM for a verdict or free-form summary, we instruct it to extract only \emph{concrete evidence facts}, map each fact to a fixed taxonomy of SSOs, and return structured records with confidence scores, line spans, and matched text. We also improve the strategy by explicitly suppressing benign metadata and generic descriptions, while still encouraging equivalence reasoning for third-party wrappers, LOTL-style invocations, and imperative setup instructions embedded in natural language. For example, given \texttt{fabric.Connection(host).run(``curl -fsSL https://x.sh | bash'')}, the extractor infers \texttt{shell\_interpreter\_execution} and \texttt{outbound\_connection}, rather than free-form results such as ``downloads and executes a script.'' 

\begin{figure}[t]
    \centering
    {
    \sffamily\fontsize{7pt}{8.4pt}\selectfont

    \begin{snapshotbox}{System Prompt}
You are extracting evidence facts, not verdicts.

\# \textbf{TASK:}
Read the artifact and extract only grounded sensitive-operation evidence.
Map each finding to the sensitive operation taxonomy using \texttt{type} and \texttt{subtype}.
Do not output benign metadata, generic capability, or mismatched intent.

\# \textbf{INPUT FORMAT:}
1. Artifact type: \{code / markdown / yaml / json / ...\}
2. Artifact text with line numbers
3. Optional context: \{package, language, file path\}

\# \textbf{OUTPUT FORMAT:}
\begin{verbatim}
{"records": [
  {
    "type": "...",
    "subtype": "...",
    "confidence": 0.0,
    "start_line": 0,
    "end_line": 0,
    "attributes": {"matched_text": "..."}
  }
]}
\end{verbatim}

\# \textbf{SPECIAL INSTRUCTIONS:}
1. Extract only concrete evidence facts grounded in the artifact text.
2. Treat third-party wrappers and LOTL/trusted-tool abuse as equivalent to native sinks.
3. If a setup instruction contains a concrete command, URL, path, or credential reference, extract it.
4. If no valid taxonomy subtype applies, return \texttt{"records": []}.
    \end{snapshotbox}
    }
    \caption{Simplified Prompt Template for the LLM-assisted Neuro Extractor.}
    \label{fig:neuro-prompt}
\end{figure}

\noindent \textbf{Neuro-to-Symbolic Feedback.} The neural extractor improves recall on semantically implicit or non-canonical SSOs, but invoking an LLM for every artifact is costly and inherently less stable than symbolic matching. To make neural discoveries reusable, we introduce a neuro-to-symbolic feedback loop that promotes stable findings into new symbolic rules. Specifically, when the neural extractor repeatedly identifies the same sensitive operation pattern across packages, such as a third-party wrapper for command execution or a recurring LOTL-style download-and-execute idiom, \toolname will send the corresponding evidence records to a dedicated \emph{specification generator}. It will abstract recurring evidence into a reusable symbolic specification by retaining the operation type, lexical trigger, argument structure, and minimal context constraints required for precise matching. The generated rule candidates are then normalized, deduplicated, and validated on a held-out corpus before being incorporated into the symbolic rule base. For example, if the neural extractor repeatedly maps \texttt{fabric.Connection(...).run(...)} to \texttt{shell\_interpreter\_execution}, the spec generator can synthesize a wrapper rule that binds \texttt{Connection.run} to the corresponding execution sink under lightweight context constraints.

\subsection{Skill Dependency Graph Generation}
\label{subsec:sdg_generation}

The second stage organizes isolated SSO records into a structured \emph{Skill Dependency Graph} (SDG) that captures how security-sensitive behaviors are grounded in artifacts, linked to their relevant operands, and connected through recovered value flow. This stage consists of three components. First, \toolname performs \emph{symbolic operand resolution} to recover the security-relevant operands of each extracted SSO and trace their explicit origins through static points-to analysis. Second, it applies \emph{LLM-assisted operand inference} to normalize semantically equivalent operands and infer missing value flow that cannot be fully recovered through symbolic analysis alone. Third, it performs \emph{SDG construction} by instantiating typed nodes and edges over artifacts, SSOs, operands, and values, thereby lifting dispersed facts into a unified graph for downstream reasoning. Notably, \toolname does not construct a complete program-level control-dependence graph. Unlike conventional programs, skill execution may depend on prompt instructions, user requests, the agent planner, and runtime context, whose conditions cannot always be mapped to precise branch semantics. 

\noindent \textbf{Symbolic Operand Resolution.} For each SSO record, we first recover the operand slots according to the operation type. 
Each major SSO category is associated with a canonical operand schema. For example, execution operations expose command-oriented operands, file operations expose path- and content-oriented operands, and network operations expose endpoint- and payload-oriented operands. We then leverage an existing points-to analysis engine~\cite{yasa-engine,wang2026yasa} to determine the source of these operands and whether multiple operations refer to the same underlying object. Specifically, \toolname traces assignments, aliases, parameter passing, return values, wrapper calls, and inter-procedural data dependencies within and across files. This allows \toolname to recover explicit operand propagation chains, such as a secret read from \texttt{os.getenv(...)} being stored in a variable and later passed into \texttt{requests.post(...)}. 

\noindent \textbf{LLM-assisted Operand Inference.} Symbolic resolution is effective when operand propagation is explicit, but real-world skills are often incomplete, weakly typed, multi-file, or mixed with prompts and instructions. As a result, some operands cannot be fully recovered through static analysis alone. To address this, we introduce an LLM-assisted operand inference component that normalizes semantically equivalent operands, infers missing value flow from local context, and bridges the gaps left by symbolic analysis. For instance, the model may infer that a variable named \texttt{token}, a prompt phrase such as ``API key,'' and a manifest field for credentials all denote the same secret-bearing operand class, even when no direct syntactic value flow is visible in the code context. 

\begin{figure}[t]
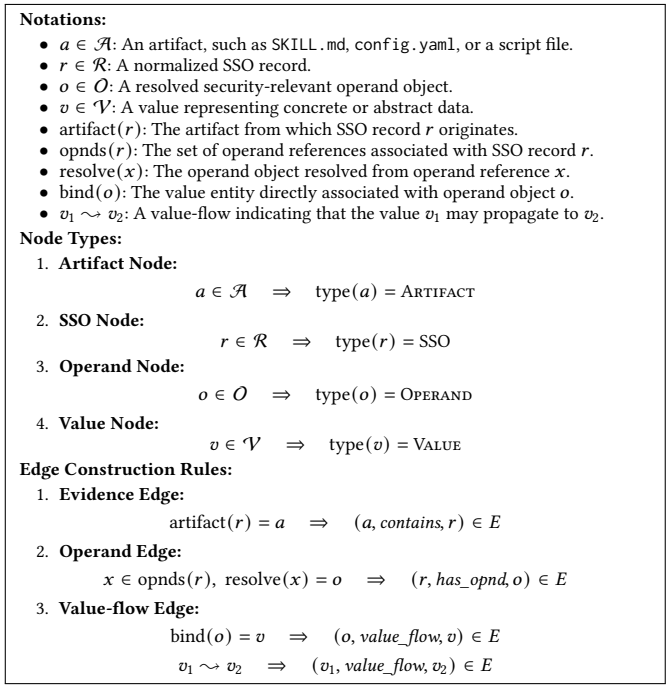

\centering
\fbox{
\begin{minipage}{0.95\linewidth}
\footnotesize
\textbf{Notations:}
\begin{itemize}[leftmargin=15pt]
    \item \(a \in \mathcal{A}\): An artifact, such as \texttt{SKILL.md}, \texttt{config.yaml}, or a script file.
    \item \(r \in \mathcal{R}\): A normalized SSO record.
    \item \(o \in \mathcal{O}\): A resolved security-relevant operand object.
    \item \(v \in \mathcal{V}\): A value representing concrete or abstract data.
    \item \(\mathrm{artifact}(r)\): The artifact from which SSO record \(r\) originates.
    \item \(\mathrm{opnds}(r)\): The set of operand references associated with SSO record \(r\).
    \item \(\mathrm{resolve}(x)\): The operand object resolved from operand reference \(x\).
    \item \(\mathrm{bind}(o)\): The value entity directly associated with operand object \(o\).
    \item \(v_1 \leadsto v_2\): A value-flow indicating that the value \(v_1\) may propagate to \(v_2\).
\end{itemize}

\textbf{Node Types:}
\begin{enumerate}[leftmargin=15pt, label=\arabic*.]
    \item \textbf{Artifact Node:}
    \[
    a \in \mathcal{A}
    \quad \Rightarrow \quad
    \mathrm{type}(a)=\textsc{Artifact}
    \]

    \item \textbf{SSO Node:}
    \[
    r \in \mathcal{R}
    \quad \Rightarrow \quad
    \mathrm{type}(r)=\textsc{SSO}
    \]

    \item \textbf{Operand Node:}
    \[
    o \in \mathcal{O}
    \quad \Rightarrow \quad
    \mathrm{type}(o)=\textsc{Operand}
    \]

    \item \textbf{Value Node:}
    \[
    v \in \mathcal{V}
    \quad \Rightarrow \quad
    \mathrm{type}(v)=\textsc{Value}
    \]
\end{enumerate}

\textbf{Edge Construction Rules:}
\begin{enumerate}[leftmargin=15pt, label=\arabic*.]
    \item \textbf{Evidence Edge:}
    \[
    \mathrm{artifact}(r)=a
    \quad \Rightarrow \quad
    (a,\textit{contains},r)\in E
    \]

    \item \textbf{Operand Edge:}
    \[
    x \in \mathrm{opnds}(r),\;\mathrm{resolve}(x)=o
    \quad \Rightarrow \quad
    (r,\textit{has\_opnd},o)\in E
    \]

    \item \textbf{Value-flow Edge:}
    \[
    \mathrm{bind}(o)=v
    \quad \Rightarrow \quad
    (o,\textit{value\_flow},v)\in E
    \]
    \[
    v_1 \leadsto v_2
    \quad \Rightarrow \quad
    (v_1,\textit{value\_flow},v_2)\in E
    \]
\end{enumerate}
\end{minipage}
}
\caption{Node and edge construction rules for the SDG.}
\label{fig:sdg-rules}
\end{figure}

\begin{figure}[t]
    \centering
    \includegraphics[width=0.98\linewidth]{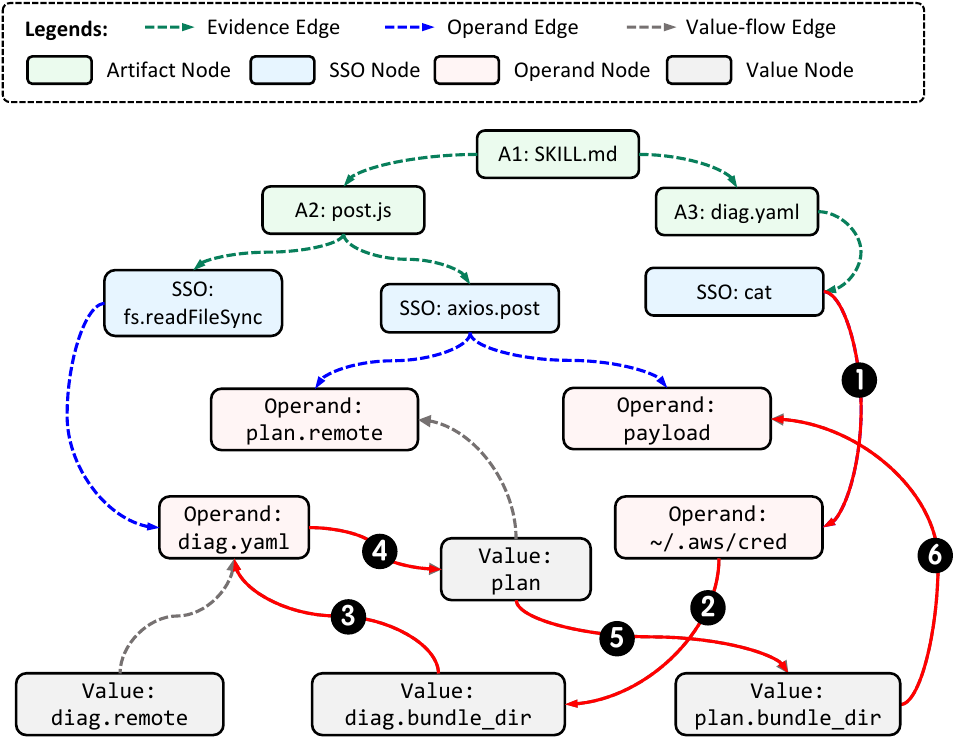}
    \caption{The SDG corresponding to the skill in ~\autoref{fig:motivation}. The \textcolor{red}{red} edge denotes the key value-flow edges for reasoning.}
    \label{fig:sdg-example}
\end{figure}

\noindent \textbf{SDG Construction.}  
After extracting SSO records and resolving their security-relevant operands, \toolname organizes them into the SDG. Formally, we model the SDG as a typed directed graph \(G=(V,E,\phi_V,\phi_E)\), where \(\phi_V\) and \(\phi_E\) assign semantic types to nodes and edges, respectively. \autoref{fig:sdg-rules} summarizes the node and edge construction rules used to instantiate the SDG.
The node set \(V\) contains four categories. \emph{Artifact} nodes represent source-level program artifacts, such as prompt files, manifests, scripts, or other analysis units. \emph{SSO} nodes represent normalized security-sensitive operations extracted by \toolname. \emph{Operand} nodes represent resolved security-relevant operand objects referenced by SSO records, such as files, endpoints, commands, or payload handles. \emph{Value} nodes represent the concrete or abstract data entities associated with operands and intermediate computations, serving as the carrier of recovered propagation semantics.
The edge set \(E\) contains three relation types. A \emph{contains} edge links an artifact node to an SSO node originating from that artifact. A \emph{has\_operand} edge links an SSO node to an operand node when that operand is explicitly referenced by the operation. A \emph{value\_flow} edge captures data propagation semantics: it either links an operand node to its directly associated value node, or links one value node to another when \toolname recovers a possible propagation relation through assignments, aliases, parameter passing, return-value binding, wrapper calls, or inter-procedural transfer. 
~\autoref{fig:sdg-example} illustrates an example SDG that shows how dispersed SSOs extracted from multiple artifacts are connected to their resolved operands and further linked through recovered value-flow dependencies. The resulting SDG lifts isolated observations into a unified context that supports subsequent reasoning over security-relevant facts.

\subsection{Neuro-Symbolic Reasoning}
\label{subsec:neuro_symbolic_reasoning}

The third stage performs malicious-skill detection through a neuro-symbolic reasoning process built on top of the SDG. 

\noindent \textbf{Pattern-based Symbolic Reasoning.}
\toolname initializes symbolic reasoning with a set of predefined patterns that capture common and security-critical malicious behaviors observed in real-world skills. Rather than relying on isolated SSO co-occurrence, each pattern is defined over connected SDG evidence and requires that the participating SSOs be semantically linked through operand dependencies, propagated values, or source-to-sink structure. At initialization, these patterns cover major classes of malicious behavior, including execution-and-delivery chains, persistence establishment, privilege escalation and identity abuse, injection and covert residency, information theft, command-and-control activity, lateral movement, defense evasion and anti-forensics, and destructive or ransomware-like actions. In essence, the symbolic component checks whether the SDG contains a behaviorally coherent subgraph whose operation sequence and dependency structure satisfy one of these known patterns. For example, in ~\autoref{fig:sdg-example}, the SDG links the secret-access operation reading \texttt{\textasciitilde/.aws/cred} to the outbound \texttt{axios.post} operation through the intermediate payload object and the value-flow edge~(highlighted in \textbf{\textcolor{red}{red}}), indicating that the sensitive content obtained from the local credential file is propagated into data sent to a remote destination.

\noindent \textbf{LLM-based Neuro Reasoning.}
Obviously, predefined patterns cannot exhaustively cover the diversity of malicious behaviors found in practice. To handle such cases, \toolname incorporates an LLM-based neuro reasoning component. Given the SDG and the symbolic matching results, the LLM is prompted with a normalized summary of the relevant subgraph, including the involved artifacts, the extracted SSOs, the resolved operands, and the value-flow paths connecting them. In this way, the neural component complements symbolic reasoning by providing semantic interpretation over incomplete, weakly matched, or previously unseen behavior patterns. When the LLM repeatedly identifies a recurring suspicious workflow that is not well covered by the current predefined patterns, \toolname records the corresponding subgraphs, operation sequences, and dependencies as candidate patterns for further use.

\section{Evaluation}

To demonstrate the effectiveness, design contribution, robustness, and ecosystem-scale practicality of \toolname, we conduct extensive experiments addressing four research questions.

\begin{itemize}[leftmargin=15pt]
    \item \textbf{RQ1: Effectiveness.}  
    How effective is \toolname when compared with state-of-the-art baselines?

    \item \textbf{RQ2: Ablation Study.}  
    How much do the major components of \toolname contribute to the overall detection performance?

    \item \textbf{RQ3: Impact of LLMs.}  
    How sensitive is \toolname to the choice of different LLMs?

    \item \textbf{RQ4: Practicality.}  
    Is \toolname practical for large-scale registries to uncover previously unknown malicious skills?
\end{itemize}

\subsection{Evaluation Setup}
\label{sec:eval-setup}

\noindent \textbf{Implementation.}
We implemented a prototype of \toolname in Python with over 7.9K lines of code, excluding third-party dependencies and external frameworks. In the parsing-based analysis, \toolname is built on top of Semgrep~\cite{semgrep}. To support realistic skill ecosystems, we incorporate a total of 2,665 Semgrep rules spanning 10 programming languages, including C, C++, C\#, Go, Java, JavaScript, PHP, Python, Ruby, and TypeScript. 
In the operand resolution, \toolname leverages the existing multi-language static analyzer YASA to perform points-to and value-flow analysis. At present, YASA supports Python, JavaScript, Java, and Go, which allows \toolname to resolve value flow and recover concrete operation objects for these languages. 
Unless otherwise specified, \toolname uses \texttt{gpt-5.3-codex-medium} as the default model in the experiments.
Due to space limitations, we do not include all prompts used in the LLM-assisted analysis in the main text. All these prompts are provided in the open-sourced artifacts.

\noindent \textbf{Running Environment.}
All experiments were conducted on a server running Ubuntu Linux 22.04.5 LTS, equipped with 256 CPUs, 1.0 TB RAM, 1 NVIDIA GeForce RTX 4090 GPU, and 28 TB of SSD storage. Unless otherwise stated, all tools were evaluated under the same environment and resource configuration.

\noindent \textbf{Dataset.}
To comprehensively evaluate \toolname, we construct two datasets, one ground truth benchmark for baseline comparison and one for large-scale in-the-wild analysis.

\textit{(1) MalSkillsBench.} 
For MalSkillsBench, we construct a labeled dataset of malicious and benign agent skills collected from real-world skill marketplaces and publicly reported malicious-skill incidents. We initially considered incorporating the malicious skills reported by MASB~\cite{liu2026maliciousagentskillswild}, which originally identified 157 malicious samples. However, a substantial portion of the corresponding GitHub repositories had become unavailable, and we were only able to successfully retrieve 57 of them. We then reran the publicly available MASB scanning pipeline on these 57 samples and found that only 6 could still be successfully detected. Moreover, these repositories are often large and artifact-heavy~(more than 4K avg.), making manual auditing and validation impractical. Since MASB is also included as one of our evaluation baselines, incorporating this small set of MASB-detected samples would additionally compromise fairness. We therefore do not include the MASB samples in the benchmark.
Instead, we construct the malicious set by systematically collecting publicly reported malicious-skill incidents from ClawHub~\cite{koi2026clawhavoc,snyk2026skills,philskents-clawhub-blocklist,opensourcemalware-clawdbot-crypto,piedpiper-openclaw-malicious-skills}. Although many of these malicious skills had already been removed from ClawHub, we recovered 597 samples from the historical commits of ClawHub's GitHub archive repository~\cite{openclaw-skills-archive}. Since many of these samples were repeatedly uploaded by the same malicious actors as part of large-scale poisoning campaigns~\cite{koi2026clawhavoc,opensourcemalware-clawdbot-crypto}, we further perform hash-based deduplication and obtain 100 unique malicious skills. Following the common practice of prior security benchmark construction~\cite{zahan2024malwarebench,cccs-cic-andmal2020,hyrum2018ember}, we build a balanced dataset by additionally selecting an equal number of popular benign skills from ClawHub. This yields a final benchmark containing 200 skills in total. For each skill, we retain its original artifact structure to preserve realistic cross-artifact analysis conditions, including prompts, manifests, scripts, configuration files, and other auxiliary artifacts.

\textit{(2) Wild-Skills-150K.}
To evaluate ecosystem-scale practicality, we further collect a large unlabeled corpus from 7 public registries and marketplaces, including 34,092 skills from Skills Directory~\cite{skillsdirectory}, 30,734 from SkillsLLM~\cite{skillsllm}, 10,598 from Smithery~\cite{smithery}, 30,789 from Skills.sh~\cite{skills.sh}, 9,643 from Skills.rest~\cite{skillsrest2026}, 11,064 from SkillsMP~\cite{skillsmp}, and 37,858 from ClawHub~\cite{clawhub,openclaw-skills-archive}. In total, this yields 164,778 raw skill entries. After deduplication, we obtain 150,108 unique skills, which we denote as \textit{Wild-Skills-150K}. This dataset is used for large-scale scanning in RQ4 to assess whether \toolname can operate efficiently and detect previously unreported malicious skills.

\noindent \textbf{Baselines.}
We compare \toolname against representative malicious skills scanning tools drawn from both industry and recent research, covering static analysis, LLM analysis, and dynamic analysis. Specifically, we include \textsc{Skill Scanner}~\cite{cisco-skill-scanner}, \textsc{Nova-Proximity}~\cite{roccia2026novaproximity}, \textsc{Skill-Sec-Scan}~\cite{huifer-skill-security-scan}, \textsc{Caterpillar}~\cite{alice2026caterpillar}, and \textsc{MASB}~\cite{liu2026maliciousagentskillswild}. To the best of our knowledge, these constitute the most representative and widely adopted publicly available end-to-end tools for malicious skill detection in the current community. We do not include Snyk's \textsc{AgentScan} in our evaluation because its detection relies on proprietary Snyk APIs and signature matching against the internal database. To ensure fair comparison, we run each baseline according to its publicly documented workflow and use its default or recommended detection settings whenever possible. For baselines that require LLM support, we use the same underlying model, \texttt{gpt-5.3-codex-medium}, across all tools. 

\subsection{RQ1: Effectiveness}
\label{sec:rq1}

We evaluate \toolname and all baselines on MalSkillsBench. For each tool, we obtain a result indicating whether a skill is malicious or benign.\footnote{For tools that output auxiliary labels such as \textit{Suspicious}, we do not count such cases as correct predictions for either the malicious or benign. As a result, the total number of predictions may not sum to the full benchmark size.} Following standard practice, we report Precision, Recall, False Positive Rate~(FPR), and F1-score.

\begin{table}[t]
\centering
\caption{Detection effectiveness on MalSkillsBench. The best results for each metric are highlighted with \textcolor{red}{red}, and the second-best results are highlighted with \textcolor{blue}{blue}. For FPR, lower values indicate better performance.}
\label{tab:rq1-main}
\resizebox{\linewidth}{!}{%
\begin{tabular}{lccrrr}
\hline
\textbf{Tool} & \textbf{TP/FP/TN/FN} & \textbf{Prec.} & \textbf{Rec.} & \textbf{FPR} & \textbf{F1} \\
\hline
\textsc{Skill Scanner~\cite{cisco-skill-scanner}} & 93/22/72/4 & \cellcolor{second}\textbf{0.81} & \cellcolor{best}\textbf{0.96} & 0.23 & \cellcolor{second}\textbf{0.88} \\
\textsc{Caterpillar}~\cite{alice2026caterpillar} & 82/40/60/18 & 0.67 & 0.82 & 0.40 & 0.74 \\
\textsc{MASB}~\cite{liu2026maliciousagentskillswild} & 9/10/90/91 & 0.47 & 0.09 & \cellcolor{second}\textbf{0.10} & 0.15 \\
\textsc{Skill-Sec-Scan}~\cite{huifer-skill-security-scan} & 40/50/50/60 & 0.44 & 0.40 & 0.50 & 0.42 \\
\textsc{Nova-Proximity}~\cite{roccia2026novaproximity} & 4/25/75/96 & 0.14 & 0.04 & 0.25 & 0.06 \\
\hline
\toolname & 92/5/95/8 & \cellcolor{best}\textbf{0.95} & \cellcolor{second}\textbf{0.92} & \cellcolor{best}\textbf{0.05} & \cellcolor{best}\textbf{0.93} \\
\hline
\end{tabular}%
}
\end{table}

\noindent \textbf{Overall Results.}
~\autoref{tab:rq1-main} presents the overall comparison results. \toolname achieves the best overall trade-off among all compared baselines, attaining the highest F1-score of 0.93 and the lowest FPR of 0.05. \toolname's recall of 0.92 is slightly lower than the best-performing \textsc{Skill Scanner}, but this small gap comes with substantially better precision and FPR. Among all baselines, the overall second-best baseline is \textsc{Skill Scanner}. As an industrial-grade scanner released by CISCO, it also leverages LLM-based analysis and achieves the highest recall, but at the cost of a much higher FPR, which leads to worse overall performance than \toolname. 
Besides, our benchmark also shows that several malicious skill detection tools perform unsatisfactorily in practice.
In particular, \textsc{Nova-Proximity} performs poorly because it mainly relies on predefined regular-expression rules, which are insufficient to capture the diverse behaviors of malicious skills. \textsc{MASB} first uses predefined rules to filter suspicious skills and then applies LLM analysis, but this simple cascade remains limited by the quality of the initial rule-based filtering stage. Even with dynamic analysis, its detection effectiveness is still unsatisfactory. This result is understandable because \textsc{MASB} was primarily designed as a large-scale empirical study of malicious skills rather than as a dedicated end-to-end detector.
Besides, \textsc{Caterpillar} and \textsc{Skill-Sec-Scan} exhibit a different trade-off. Both tools can detect a non-trivial portion of malicious skills, but they sacrifice precision and false positive control. 
Overall, \toolname provides a better balance between precision and recall.

\noindent \textbf{FP/FN Analysis.}
We conduct a further review of the FPs and FNs. The FPs are concentrated in skills whose descriptions contain strong operational or automation-related signals, such as installation instructions, environment configuration steps, cron/heartbeat workflows, and web scraping examples. Although such content is benign in context, it partially overlaps with the malicious patterns that \toolname treats as security-relevant evidence. 
On the other side, the FNs mainly arise in cases where maliciousness is conveyed more implicitly than our current SSO categories capture. Representative examples include documentation-centric social engineering, guidance that redirects users to external downloads, and bot-control capabilities. In these cases, the malicious signal is not always expressed as an explicit sensitive operation, but instead emerges through intent, indirection, or downstream enablement. This suggests that the main limitation of \toolname lies in covering more implicit and document-oriented threat manifestations.

\subsection{RQ2: Ablation Study}
\label{sec:rq2}

We conduct an ablation study on MalSkillsBench using the same evaluation setting as in RQ1. To understand the contribution of each major component, we evaluate the following variants:

\begin{itemize}[leftmargin=15pt]
    \item \textit{Full \toolname.} The complete system with both symbolic and neural extraction, followed by symbolic and neural reasoning.

    \item \textit{w/o Symbolic Extractor.} This variant removes the symbolic extraction component and relies only on the neural extractor to identify SSO from the skill.

    \item \textit{w/o Neuro Extractor.} This variant removes the neural extractor for SSO collection.

    \item \textit{w/o Symbolic Reasoner.} This variant removes the symbolic reasoning and classifies based only on neural reasoning.

    \item \textit{w/o Neuro Reasoner.} This variant removes the neural reasoning and relies only on symbolic reasoning.
\end{itemize}

\begin{table}[t]
\centering
\caption{Ablation study of \toolname with 4 variants. The best results for each metric are highlighted with \textcolor{red}{red}. For FPR, lower values indicate better performance.}
\label{tab:rq2-ablation}
\resizebox{\linewidth}{!}{%
\begin{tabular}{lccccc}
\hline
\textbf{Variant} & \textbf{TP/FP/TN/FN} & \textbf{Prec.} & \textbf{Rec.} & \textbf{FPR} & \textbf{F1} \\
\hline
\textit{Full \toolname} & 92/5/95/8 & \textbf{0.95} & \cellcolor{best}\textbf{0.92} & \textbf{0.05} & \cellcolor{best}\textbf{0.93} \\
\hline
\textit{w/o Symbolic Extractor} & 27/0/100/73 & \cellcolor{best}\textbf{1.00} & 0.27 & \cellcolor{best}\textbf{0.00} & 0.43 \\
\textit{w/o Neuro Extractor} & 91/5/95/9 & \textbf{0.95} & 0.91 & \textbf{0.05} & \cellcolor{best}\textbf{0.93} \\
\textit{w/o Symbolic Reasoner} & 92/5/95/8 & \textbf{0.95} & \cellcolor{best}\textbf{0.92} & \textbf{0.05} & \cellcolor{best}\textbf{0.93} \\
\textit{w/o Neuro Reasoner} & 82/49/51/18 & 0.63 & 0.82 & 0.49 & 0.71 \\
\hline
\end{tabular}%
}
\end{table}

\autoref{tab:rq2-ablation} presents the ablation results. Overall, the full \toolname achieves the best overall trade-off between effectiveness and FP control, with an F1-score of 0.93 and an FPR of 0.05. 

\noindent \textbf{\textit{w/o Symbolic Extractor.}} Removing the symbolic extractor causes \toolname to fail in a highly conservative manner. Although this variant achieves perfect precision (1.00) and zero FPR, its recall drops sharply from 0.92 to 0.27, indicating that it becomes overly conservative and misses most malicious skills. This suggests that the neural extractor alone identifies only a narrow subset of highly salient SSOs, while failing to robustly recover the broader range of suspicious operations needed for high recall. One reason is that the system prompt used by the neural extractor is primarily designed to uncover stealthy, implicit, or adversarial SSOs, rather than to exhaustively extract all generic SSOs from a skill. As a result, when used in isolation, the neural extractor tends to be conservative. 

\noindent \textbf{\textit{w/o Neuro Extractor.}} In contrast, removing the neuro extractor causes almost no performance drop on MalSkillsBench. This suggests that, for the current benchmark, the symbolic extractor already covers most of the SSOs needed to identify known malicious skills. This result is reasonable given how the benchmark is constructed. Since the ground-truth malicious samples are collected from already discovered attacks, many of their malicious behaviors are relatively explicit and can be captured by static rules. Nevertheless, this should not be interpreted as evidence that the neuro extractor is unnecessary. Its role is to complement symbolic extraction by capturing stealthier, more implicit, or adversarial SSOs that may be underrepresented in the current benchmark but are likely to become increasingly important as the skills ecosystem evolves.

\noindent\textbf{\textit{w/o Neuro Reasoner.}} Removing the neuro reasoning component causes the largest overall degradation, reducing F1 from 0.93 to 0.71 and increasing FPR from 0.05 to 0.49. This shows that symbolic reasoning alone is insufficient for reliably distinguishing malicious from benign skills at the semantic level. In particular, although symbolic reasoning can capture suspicious operations, it is less effective at interpreting intent and context. For example, a pattern such as \texttt{curl | bash} may appear suspicious structurally, but its actual risk depends on contextual factors such as the source being fetched. Therefore, neuro reasoning is the key component for controlling FPs and making accurate decisions.

\noindent\textbf{\textit{w/o Symbolic Reasoner.}} Finally, removing symbolic reasoning does not change the detection metrics on this benchmark, suggesting that neuro reasoning dominates the final classification performance in this setting. However, the main purpose of symbolic reasoning is not necessarily to improve benchmark accuracy in isolation. Instead, it provides structured intermediate constraints that can guide downstream reasoning, reduce unnecessary LLM involvement, and improve cost-effectiveness and reasoning stability in large-scale deployment. In our design, symbolic reasoning serves as an efficient first-stage filter. Although it may still produce FPs or miss some cases when used alone, these limitations can be mitigated through neuro-symbolic feedback to improve recall, while the subsequent neuro reasoning further suppresses FPs. Therefore, the symbolic-then-neural reasoning strategy is important not only for effectiveness, but also for scalable ecosystem-wide scanning.

% \autoref{tab:rq2-ablation} presents the ablation results. Overall, the full version of \toolname achieves the best performance, with an F1-score of 0.93 and an FPR of 0.05. Among all ablations, removing the neuro reasoning component causes the largest performance drop, reducing F1 to 0.71 and increasing FPR to 0.49. This indicates that neuro reasoning is critical for accurately distinguishing malicious skills from benign ones while controlling false positives.
% Removing the symbolic extractor also substantially hurts recall, which drops from 0.92 to 0.27, even though precision becomes perfect and FPR falls to 0.00. This suggests that the symbolic extractor contributes essential high-coverage evidence, and without it the system becomes overly conservative, identifying only a small subset of malicious skills.
% In contrast, removing the neuro extractor leads to only a minor performance decrease, while removing symbolic reasoning yields no observable degradation on this benchmark. This is consistent with the intended roles of these two components. The neuro extractor is mainly designed to capture more stealthy and adversarial SSOs, and such benefits may not be fully reflected on the current benchmark. Meanwhile, symbolic reasoning is primarily intended to leverage structured intermediate evidence to constrain the reasoning process and reduce unnecessary token consumption, improving efficiency and reasoning stability rather than directly boosting detection metrics in every setting.

\subsection{RQ3: Impact of LLMs}
\label{sec:rq3}

\begin{table}[t]
\centering
\caption{Performance of \toolname with different LLMs.}
\label{tab:rq3-llm}
\resizebox{\linewidth}{!}{%
\begin{tabular}{lccccc}
\hline
\textbf{LLM} & \textbf{TP/FP/TN/FN} & \textbf{Prec.} & \textbf{Rec.} & \textbf{FPR} & \textbf{F1} \\
\hline
\textit{GPT-5.3-CodeX-Medium} & 92/5/95/8 & \cellcolor{best}\textbf{0.95} & \cellcolor{second}\textbf{0.92} & \cellcolor{best}\textbf{0.05} & \cellcolor{best}\textbf{0.93} \\
\hline
\textit{Claude-Sonnet-4.6} & 70/14/86/30 & 0.83 & 0.70 & 0.14 & 0.76 \\
\textit{Gemini-3.1-Flash-Lite} & 95/24/76/5 & 0.80 & \cellcolor{best}\textbf{0.95} & 0.24 & \cellcolor{second}\textbf{0.87} \\
\textit{Qwen3.5-397B-A17B} & 78/15/85/22 & 0.84 & 0.78 & 0.15 & 0.81 \\
\textit{DeepSeek-V3.2} & 71/5/95/29 & \cellcolor{second}\textbf{0.93} & 0.71 & \cellcolor{best}\textbf{0.05} & 0.81 \\
\hline
\end{tabular}%
}
\end{table}

To understand how the choice of LLM affects \toolname, we run \toolname with several widely used LLMs under the same experimental setting, including Claude Sonnet 4.6, Gemini 3.1 Flash-Lite, Qwen3.5-397B-A17B, and DeepSeek-V3.2. \autoref{tab:rq3-llm} reports the results.
Overall, the choice of LLMs has a clear impact on the effectiveness of \toolname, mainly by shifting the trade-off between recall and FP control. GPT-5.3-CodeX-Medium achieves the best overall performance, obtaining the highest precision (0.95), the lowest FPR (0.05, tied), and the best F1-score (0.93). This suggests that it provides the most balanced behavior across both extraction and reasoning, enabling \toolname to identify malicious skills accurately without introducing excessive false alarms.
Gemini-3.1-Flash-Lite achieves the highest recall (0.95), correctly identifying almost all malicious skills, but this comes at the cost of a noticeably higher FPR (0.24). This pattern suggests that Gemini tends to behave more aggressively, surfacing a larger set of potentially suspicious behaviors and therefore missing fewer malicious samples, but at the same time over-classifying more benign skills as malicious.
DeepSeek-V3.2 exhibits the opposite behavior. It achieves very high precision (0.93) and the lowest FPR (0.05, tied with GPT), but its recall drops to 0.71. This indicates a more conservative decision boundary. Claude Sonnet 4.6 shows a similar but less favorable pattern, with both lower recall and lower overall F1. Qwen3.5-397B-A17B provides a more balanced middle ground, although it still falls short of GPT in both recall and FP control.
These results indicate that \toolname is robust across different model families, but the quality of the underlying LLM still matters. Stronger or more code- and instruction-reliable models can noticeably improve the end-to-end effectiveness.

% \subsection{RQ4: Practicality}
% \label{sec:rq4}

% To evaluate the real-world practicality, we run \toolname on \textit{Wild-Skills-150K}, which contains 150,108 deduplicated skills collected from 7 public registries and marketplaces. In total, \toolname reported 620 skills as malicious. Among them, 97 were from Skills Directory, 51 from SkillsLLM, 39 from Smithery, 11 from Skills.sh, 112 from Skills.rest, 58 from SkillsMP, and 252 from ClawHub.
% To assess the quality of these findings, we randomly sampled 100 flagged skills for manual review. The first two authors independently examined each sample based on its full artifact set. For cases where the two reviewers disagreed, a third researcher was introduced to arbitrate and determine the final label. Through this process, we confirmed that 76 of the 100 sampled skills are previously unreported malicious samples. These confirmed cases exhibit diverse malicious behaviors, including credential theft, suspicious download-and-execute logic, and malicious dependency installation.
% All these 76 malicious skills confirmed through manual review have been responsibly reported to the corresponding platforms and maintainers, and are currently awaiting their responses. The remaining flagged samples are also under continuous review. Overall, this large-scale study demonstrates that \toolname is practical for ecosystem-scale scanning and capable of uncovering previously unknown malicious skills in real-world skill registries.

\subsection{RQ4: Practicality}
\label{sec:rq4}

\begin{table}[t]
\centering
\caption{Distribution of the originally reported and manually confirmed malicious skills.}
\label{tab:rq4-wildskills}
\resizebox{\linewidth}{!}{%
\begin{tabular}{lr|rrrrrr}
\hline
\textbf{Source} & \textbf{Orig./Conf.} & \textbf{P1} & \textbf{P2} & \textbf{P3} & \textbf{P4} & \textbf{P5} & \textbf{P6} \\
\hline
ClawHub~\cite{clawhub}          & 252/230 & 204 & 16 & 8 & - & 1 & 1 \\
Skills Directory~\cite{skillsdirectory} & 97/25   & 20  & 3  & 1 & 1 & - & - \\
Skills.sh~\cite{skills.sh}        & 11/7    & 7   & -  & - & - & - & - \\
SkillsLLM~\cite{skillsllm}        & 51/25   & 24  & 1  & - & - & - & - \\
SkillsMP~\cite{skillsmp}         & 58/35   & 32  & 2  & 1 & - & - & - \\
Skills.rest~\cite{skillsrest2026}      & 112/58  & 54  & 3  & 1 & - & - & - \\
Smithery~\cite{smithery}         & 39/20   & 18  & 1  & 1 & - & - & - \\
\hline
\textbf{Total}   & \textbf{620/400} & \textbf{359} & \textbf{26} & \textbf{12} & \textbf{1} & \textbf{1} & \textbf{1} \\
\hline
\end{tabular}%
}
\vspace{0.5ex}

\raggedright
\footnotesize
\textbf{Pattern Labels:}
P1: Data Theft/Exfiltration;
P2: Instruction Override/Deception;
P3: Remote Code Execution;
P4: Remote Fetch/Supply Chain;
P5: Privilege Escalation/Lateral Movement;
P6: Stealth/Evasion/Obfuscation.
\end{table}

\noindent \textbf{Overall Results.} To evaluate the real-world practicality of \toolname, we applied it to \textit{Wild-Skills-150K}. \toolname initially flagged 620 potentially malicious skills. We then manually reviewed all reported cases. Specifically, the first two authors, who have 4 and 2 years of security experience, independently examined each reported skill and determined whether it exhibited malicious behavior. For cases with inconsistent conclusions, a third author with 7 years of security experience was introduced to arbitrate and assign the final label.
The manual review confirms that 400 of the 620 flagged skills are indeed malicious. The resulting precision is lower than that on MalSkillsBench because RQ1 uses a balanced benchmark, whereas RQ4 operates in an open-world setting where malicious skills are rare. Although the FPR cannot be computed because TN remains unknown, the remaining 220 flagged skills correspond to nearly 0.15\% of all 150,108 scanned skills. Many of these cases arise from benign skills that contain risky-looking patterns, such as remote installation behaviors (e.g., \texttt{curl | bash}) or credentialed access to remote endpoints. Determining whether such behaviors are truly malicious often requires inspecting second-stage payloads or richer execution contexts, and thus manual review remains necessary. As of now, all confirmed malicious skills have been responsibly reported and are currently awaiting confirmation from the platforms.

\noindent \textbf{Registry Distribution.} As shown in \autoref{tab:rq4-wildskills}, the confirmed malicious skills are distributed across all 7 registries, with ClawHub contributing the largest portion. Notably, even registries that claim to incorporate security mechanisms still contained malicious skills. For example, ClawHub is protected by its own moderation engine~\cite{clawhub-moderation-engine} together with VirusTotal~\cite{openclaw-virustotal}, while Skills.sh~\cite{skills-sh-audits} integrates Gen Agent Trust Hub~\cite{gen-skill-scanner}, Socket~\cite{socket-skills-security}, and Snyk~\cite{snyk-vercel-skill-security}, and Skills Directory also claims security checks~\cite{skillsdirectory-security}. Despite these protections, we still identified 262 previously undisclosed malicious skills in these three registries. These findings suggest that currently deployed scanning mechanisms are still insufficient to reliably prevent malicious skills from entering public ecosystems.

\noindent \textbf{Malicious Pattern.} In terms of attack behaviors, data theft and exfiltration dominate the confirmed malicious skills~(89.8\%), followed by instruction override and deception, remote code execution, remote fetch or supply-chain abuse, privilege escalation, and stealth or obfuscation.
% These results reveal that malicious skills are not confined to a single behavior type, but already cover multiple attack strategies with different operational goals.
Overall, these findings demonstrate that \toolname is practical for ecosystem-scale scanning and effective at uncovering previously unknown malicious skills in real-world registries.
\section{Discussion}
\noindent \textbf{Limitations.}
Our study has several limitations. First, although \textsc{MalSkillsBench} is constructed from real-world incidents, it may not fully cover the diversity of malicious skills in the wild. 
% Publicly reported and archived cases are inherently biased toward attacks that were discovered and preserved. 
That said, we believe it still provides a realistic and reproducible foundation for comparison, especially given the current lack of public benchmarks.
Second, while \toolname supports multi-language parsing, deeper operand resolution currently depends on YASA~\cite{yasa-engine,wang2026yasa} and is limited to Python, JavaScript, Java, and Go. 
% As a result, value-flow analysis may be less useful in other languages. 
Nevertheless, this limitation is not fundamental, and future works can incorporate additional analyzers to improve the prototype.
Third, as malicious-skill ecosystems evolve, attackers may adopt more implicit or adversarial strategies that reduce explicit security signals. Although \toolname is intentionally designed to avoid direct maliciousness judgment and instead reason over extracted SSOs, it may still be affected by AI-evasion strategies~\cite{checkpoint2025aievasion,melo2025aimalwarepromptinjection}. More broadly, malicious-skill detection remains an ongoing hide-and-seek between attackers and defenders, and we therefore view this work as a foundation for future research rather than a once-and-for-all solution.

\noindent \textbf{Implications.}
Our findings suggest three broader implications for securing the emerging agentic supply chain. First, defenses for traditional software supply chains cannot be directly transferred to skill ecosystems. Unlike traditional packages, the risk of a skill often emerges from interactions among heterogeneous artifacts, so the defenses should treat cross-artifact reasoning as a basic requirement.
Second, our results indicate that current protection mechanisms remain insufficient in practice. Although some registries have already integrated protections and the community has developed a range of scanning tools, our benchmark results and real-world findings show that these mechanisms still miss malicious skills or incur substantial false positives. 
Third, we believe that both the \textsc{MalSkillsBench} and the neuro-symbolic design of \toolname can serve as a foundation for future research and inspire more effective defenses for these evolving threats.

\section{Related Work}
\label{sec:related_work}

\noindent \textbf{OSS Supply Chain Security.}
Open-source software (OSS) supply chain security has become a major research focus as attackers increasingly target package registries and dependency ecosystems. Prior works~\cite{gu2023package,ladisa2023sokattack} have systematically studied the attack surface of traditional OSS supply chains, including typosquatting~\cite{ladisa2023sokattack,xie2025squatting}, dependency confusion~\cite{ladisa2023sokattack}, and even package confusion~\cite{neupane2023typosquatting}. In response, researchers have proposed a wide range of detection and mitigation approaches. \emph{Traditional methods} detect suspicious behaviors through semantic rules~\cite{li2023malwukong,zheng2024oscar}, taint-style analysis~\cite{li2023malwukong,duan2021maloss}, and sandbox execution~\cite{zheng2024oscar,duan2021maloss}. \emph{Learning-based methods} use metadata~\cite{sajal2024metadata} and source-code features~\cite{sejfia2022practical,ladisa2023feasibility} to train classifiers. More recently, \emph{LLM-based approaches} have been explored for sensitive API identification~\cite{huang2024spiderscan}, data augmentation~\cite{gao2024malguard}, code understanding~\cite{wang2025malpacdetector,yu2024maltracker}, and maliciousness judgment~\cite{zahan2025socketai,di2024posterllm}. In contrast, \toolname targets agent skills, whose malicious behavior is often distributed across artifacts rather than code scripts only.

\noindent \textbf{Agentic Supply Chain Security.}
With the rise of LLM agents, recent work has started to examine security risks in the agentic supply chain~\cite{hu2025llmsc,wang2025llmsc,huang2025llmsc}, including agent skills~\cite{liu2026maliciousagentskillswild,liu2026vulskill,guo2026skillprobesecurityauditingemerging} and MCP servers~\cite{hou2026mcp,wang2025mcpguardautomaticallydetecting}. Public reports and benchmark efforts have shown that real-world skill ecosystems already contain attacks such as wallet theft~\cite{koi2026clawhavoc,opensourcemalware-clawdbot-crypto} and reverse shells~\cite{snyk2026skills}. Preliminary defenses include rule-based scanners~\cite{huifer-skill-security-scan,goplus-agentguard}, LLM-based analysis~\cite{cisco-skill-scanner}, and dynamic analysis~\cite{liu2026maliciousagentskillswild} for skills. However, most existing tools analyze individual artifacts and provide limited support for cross-artifact reasoning. \toolname addresses this gap by focusing on neuro-symbolic reasoning on heterogeneous artifacts.

\noindent \textbf{Agent Security.}
Recent work has shown that LLM-based agents are highly vulnerable to taint-style and prompt-driven attacks. In particular, prior studies have demonstrated that attackers can manipulate agent inputs to steer model outputs into generating malicious code~\cite{liu2024llmrce} or SQL queries~\cite{pedro2025prompt2sql}, which leads to a series of injection vulnerabilities. In response, recent efforts have proposed analysis techniques for detecting such vulnerabilities, including static taint analysis~\cite{icse26taintp2x} and greybox fuzzing~\cite{liu2024llmrce,liu2025agentfuzz}. In comparison, our work considers a complementary threat model in which adversaries inject malicious skills into the agent ecosystem and compromise system security.
\section{Conclusion}
In this paper, we presented \toolname, a neuro-symbolic framework for detecting malicious skills in the emerging agentic supply chain. By combining cross-artifact SSO extraction, skill dependency graph construction, and neuro-symbolic reasoning, \toolname is designed to better handle the heterogeneous, context-dependent, and adversarial nature of malicious skills.
Our evaluation on MalSkillsBench shows that \toolname achieves the best overall detection performance among representative baselines, while our large-scale deployment on 7 public registries further demonstrates its practicality and ability to uncover previously unknown malicious skills in the wild. Overall, we hope this work can serve as a foundation for future research on securing the agentic supply chain.

\section*{Acknowledgements}
We used GenAI tools for language refinement. We reviewed all AI-assisted content and take full responsibility for this paper. This work was supported in part by the National Natural Science Foundation of China (grants No.62572209, 62502168), the Hubei Provincial Key Research and Development Program (grant No. 2025BAB057), and Ant Group Research Fund.

\section*{Data Availability Statement}
We have made the implementations of \toolname and experimental data publicly accessible at \url{https://github.com/security-pride/MalSkills}.

\balance
\bibliographystyle{ACM-Reference-Format}
\bibliography{reference}

\end{document}